\documentclass[sigconf]{acmart}
\AtBeginDocument{%
  \providecommand\BibTeX{{%
    Bib\TeX}}}

\setcopyright{none}
\renewcommand\footnotetextcopyrightpermission[1]{}
\acmDOI{}
\acmISBN{}

\usepackage{amsfonts}
\usepackage{booktabs}
\usepackage{algorithmic}
\usepackage{xcolor}
\usepackage{graphicx}
\usepackage{adjustbox}
\usepackage{appendix}
\usepackage[inline]{enumitem}
\usepackage{textcomp}
\usepackage{lipsum}
\usepackage{afterpage}
\usepackage{marvosym}
\usepackage{tikz}
\usepackage{float}
\usepackage{placeins}
\usepackage{listings}
\usepackage{pifont}
\usepackage{cleveref}
\usepackage{xargs}
\usepackage{tabularray}
\usepackage{longtable}
\usepackage{booktabs}
\usepackage{tabularx}
\usepackage{array}
\newcolumntype{Y}{>{\raggedright\arraybackslash}X}

\usepackage[T1]{fontenc}

\definecolor{KeywordBlue}{RGB}{0,92,197}
\definecolor{OpRed}{RGB}{178,34,34}
\definecolor{NumTeal}{RGB}{0,128,128}

\crefname{listing}{Listing}{Listings} 
\lstdefinestyle{pyopendt}{
  language=Python,
  basicstyle=\ttfamily\small,
  frame=single,
  breaklines=true,
  breakatwhitespace=true,
  showstringspaces=false,
  columns=fullflexible,
  stringstyle=\color{black},
  commentstyle=\color{CommentGray},
  keywordstyle=\color{KeywordBlue},
  emph={json,parser,sim_results,batch_data,current_topology,slo_targets},
  emphstyle=\color{KeywordBlue},
  literate=
   *{+}{{{\color{OpRed}+}}}1
    {-}{{{\color{OpRed}-}}}1
    {*}{{{\color{OpRed}*}}}1
    {/}{{{\color{OpRed}/}}}1
    {=}{{{\color{OpRed}=}}}1
    {:}{{{\color{OpRed}:}}}1
    {,}{{{\color{OpRed},}}}1
    {(}{{{\color{OpRed}(}}}1
    {)}{{{\color{OpRed})}}}1
    {[}{{{\color{OpRed}[}}}1
    {]}{{{\color{OpRed}]}}}1
    {.}{{{\color{OpRed}.}}}1
    {0}{{{\color{NumTeal}0}}}1
    {1}{{{\color{NumTeal}1}}}1
    {2}{{{\color{NumTeal}2}}}1
    {3}{{{\color{NumTeal}3}}}1
    {4}{{{\color{NumTeal}4}}}1
    {5}{{{\color{NumTeal}5}}}1
    {6}{{{\color{NumTeal}6}}}1
    {7}{{{\color{NumTeal}7}}}1
    {8}{{{\color{NumTeal}8}}}1
    {9}{{{\color{NumTeal}9}}}1
}

\newcommandx{\todolarge}[2][1=]{\todo[inline,size=\large,linecolor=blue,backgroundcolor=cyan,bordercolor=blue,#1]{#2}}

\definecolor{darkred}{rgb}{0.5,0,0}
\definecolor{darkgreen}{rgb}{0,0.5,0}
\definecolor{darkblue}{rgb}{0,0,0.5}

\newcommand{\circled}[1]{%
    \tikz[baseline=(char.base)]{
        \node[shape=circle, fill=black, text=white, inner sep=0.5pt, font=\small] (char) {#1};
    }%
}

\def\BibTeX{{\rm B\kern-.05em{\sc i\kern-.025em b}\kern-.08em
    T\kern-.1667em\lower.7ex\hbox{E}\kern-.125emX}}


\newcommand{\ourtool}{OpenDT}
\newcommand{\oursim}{OpenDC}
\newcommand{\footprinter}{FootPrinter}
\begin{document}

\title{OpenDT: Exploring Datacenter Performance and \\ Sustainability with a Self-Calibrating Digital Twin \\ \LARGE{\textcolor{gray}{[Technical Report on the ICPE HCP homonym article]}}}

\author{Radu Nicolae}
\email{R.Nicolae@vu.nl}
\orcid{0009-0007-0318-9266}
\affiliation{%
  \institution{Vrije Universiteit Amsterdam}
  \city{Amsterdam}
  \country{The Netherlands}
}

\author{Jules van der Toorn}
\email{J.vander.Toorn@student.vu.nl}
\orcid{0000-0001-8731-9783}
\affiliation{
  \institution{Vrije Universiteit Amsterdam}
  \city{Amsterdam}
  \country{The Netherlands}
}

\author{Stavriana Kraniti}
\email{S.Kraniti@student.vu.nl}
\orcid{0009-0002-2248-2736}
\affiliation{%
  \institution{Vrije Universiteit Amsterdam}
  \city{Amsterdam}
  \country{The Netherlands}
}

\author{Houcen Liu}
\email{H.Liu19@student.vu.nl}
\orcid{0009-0001-4693-0743}
\affiliation{%
  \institution{Vrije Universiteit Amsterdam}
  \city{Amsterdam}
  \country{The Netherlands}
}

\author{Alexandru Iosup}
\email{A.Iosup@vu.nl}
\orcid{0000-0001-8030-9398}
\affiliation{%
  \institution{Vrije Universiteit Amsterdam}
  \city{Amsterdam}
  \country{The Netherlands}
}

\thanks{$^{*}$First four authors contributed equally to this research. \\ $^{+}$The fifth author adds conceptual contribution and supervision.}  

\renewcommand{\shortauthors}{Radu Nicolae, Jules van der Toorn, Stavriana Kraniti, Houcen Liu, \& Alexandru Iosup}
\renewcommand{\shorttitle}{OpenDT:
Exploring Datacenter Ops with a Self-Calibrating Digital Twin}

\begin{abstract}
Datacenters are the backbone of our digital society, but raise numerous operational challenges.
We envision digital twins becoming primary instruments in datacenter operations, continuously and autonomously helping with major operational decisions and with adapting ICT infrastructure, live, with a human-in-the-loop.
Although fields such as aviation and autonomous driving successfully employ digital twins, an open-source digital twin for datacenters has not been demonstrated to the community.
Addressing this challenge, we design, implement, and experiment using \underline{\ourtool{}}, an \underline{Open}-source, \underline{D}igital \underline{T}win for monitoring and operating datacenters through a continuous integration cycle that includes: 
(1) live and continuous telemetry data;
(2) discrete-event simulation using live telemetry from the physical ICT, with self-calibration; and
(3) SLO-aware and human-approved feedback to physical ICT. 
Through trace-driven experiments with a prototype mainly covering stages~1 and~2 of the cycle, we show that (i)~\ourtool{} can be used to reproduce peer-reviewed experiments and extend the analysis with performance and energy-efficiency results; (ii)~\ourtool{}'s online re-calibration can increase digital-twinning accuracy, quantified to a MAPE 4.39\% vs. 7.86\% in peer-reviewed work. 
\ourtool{} adheres to 
FAIR/FOSS
principles and is available at: 
\url{https://github.com/atlarge-research/opendt/tree/hcp}. 
\end{abstract}



\maketitle

\section{Introduction} \label{\ourtool{}:sec:intro}

Our digitalized society and economy increasingly rely on digital services running in increasingly larger datacenters~\cite{DBLP:journals/corr/abs-2206-03259, DBLP:conf/sc/AndreadisVMI18, market:IDC24AI}. 
Operators use system analysis, increasingly based on simulation, for designing and operating datacenters~\cite{DBLP:journals/corr/abs-2206-03259, nicolae5377101m3sa, DBLP:conf/ccgrid/MastenbroekAJLB21}. 
Using simulation as the core of digital twins that mimic datacenter conditions, and support analysis and decision-making in datacenter operations, is an open, emerging challenge~\cite{DBLP:conf/compsac/MolanKBTPCIRRVP23}. 
In this work, we design, prototype, and experiment using \underline{\ourtool{}}, an \underline{Open}-source \underline{D}igital \underline{T}win for datacenters.

\begin{figure}[t]
    \centering
    \vspace*{0.2cm}
    \includegraphics[width=0.98\linewidth]{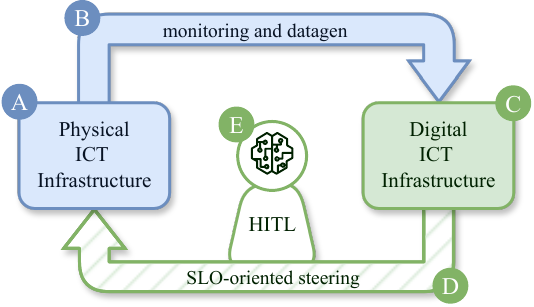}
    \vspace*{-0.2cm}
    \caption{High-level overview of datacenter digital twinning: 
    (1) The Physical ICT infrastructure collects telemetry data, 
    (2) the Digital ICT infrastructure ingests telemetry and related data, then 
    (3) mimics (twins) the operation of the Physical ICT, 
    and (4) offers back SLO-aware adjustment feedback. 
    A human-in-the-loop (HITL) oversees the process.
    }
    \label{fig:high-level-element}
    \vspace*{-0.5cm}
\end{figure}

Simulation already supports datacenter operators and scientists with timely and cost-efficient experimentation and analysis~\cite{Iosup2024DigitalTwins, DBLP:conf/ccgrid/MastenbroekAJLB21}. 
For example, in the European Horizon project Graph Massivizer, simulators predict speedup, failure cost, energy consumption, and CO2-emissions for massive-scale infrastructure~\cite{nicolae5377101m3sa, DBLP:conf/compsac/MolanKBTPCIRRVP23, DBLP:conf/wosp/IosupPVTMHZBFK23, DBLP:conf/wosp/Sanchez0RP23}.

Although simulators are valuable for detailed, realistic analysis of datacenters under workload, in ICT, there is currently no closed-loop process that 
continuously ingests live telemetry, updates the state inside the simulation, 
and calibrates the simulator so the multi-metric analysis matches reality.
Instead, these steps must be taken independently, with considerable delays and potential errors particularly at boundaries.

\Cref{fig:high-level-element} illustrates the \textit{digital twinning process} that informed the high-level design of \ourtool{}. 
This process involves continuous communication between a physical twin~(label \circled{A}) and a digital twin~(\circled{C}). 
The physical twin generates telemetry data~(\circled{B}), either obtained from the ICT infrastructure's monitoring and logging systems, or generated from it via advanced analytics or AI-based processes; telemetry data is ingested into the digital twin to be processed directly, or placed in a larger data pool. The digital twin uses techniques to mimic (\textit{twin}) the operation of the ICT infrastructure, often simulation-based and leveraging SLO-oriented analysis. Closing the loop, the digital twin informs and potentially steers the physical datacenter with SLO-oriented decisions and recommendations. 
A \textit{human-in-the-loop}~(\circled{E}), aiming to ensure the correct and ethical operation of the process, oversees at least the major decisions and can intervene at any time.

In this work, with \ourtool{}, we make three main contributions:

\begin{enumerate}[label=\textbf{C\arabic*.},align=left,leftmargin=*,widest={9},topsep=3pt,itemindent=12pt]
\item \label{introduction:c1} \textbf{Design and prototype}: 
We design and prototype \ourtool{} as a datacenter digital twin.
The current version of \ourtool{} supports the cycle of continuous digital-twinning through: 
(1) telemetry ingestion, 
(2) trace- and configuration-driven, discrete-event, SLO-aware simulation for multi-metric datacenter analysis,
(3) support for human-in-the-loop digital twinning.
\ourtool{} is in its early design and prototype stages, with future work in all its main elements, in particular, closing the twinning loop with automated steering~(\Cref{fig:high-level-element}, \circled{D}). 

\item \label{introduction:c2} \textbf{Exploration}: 
We validate \ourtool{} through real-world experimentation using its prototype. 
We reproduce and then extend a peer-reviewed experiment~\cite{DBLP:conf/wosp/NiewenhuisTIM24}. We explore how much \ourtool{}'s self-calibration improves accuracy.
All experiments use real-world workload and energy traces from SURF, the Dutch national supercomputing center~\cite{DBLP:journals/fgcs/VersluisCGLPCUI23}. 

\item \label{introduction:c3} \textbf{Open Science}: We follow the principles of FAIR, FOSS, and reproducible science, and release the engineered \ourtool{} prototype, together with a reproducibility capsule at:\\ \url{https://github.com/atlarge-research/opendt/tree/hcp} .

\end{enumerate}
\section{Design of \ourtool{}: an operational ecosystem for datacenter digital twinning}\label{sec:design}

In this section, we first identify design requirements based on datacenter operational needs, then present the architecture of \ourtool{}.

\subsection{Requirement and use case analysis}
We identify three main stakeholders, within datacenter operation, computer systems research, and education, which we describe in use case 1 (UC1), UC2, and UC3, respectively:

\label{sec:design:usecases}\label{sec:design:reqs}\label{sec:design:requirements}\label{sec:design:usecases}

\begin{enumerate}[label=\textbf{(UC\arabic*)},leftmargin=0pt,itemindent=3em]
    \item \label{design:uc1:practitioner} \label{design:uc1} \textbf{Monitoring and operation datacenters}: 
    OpenDT should aid practitioners in (1) monitoring the performance and sustainability of datacenters and (2) conducting infrastructure steering decisions, validated through simulation. OpenDT should ensure high prediction accuracy \ref{design:nfr:accuracy}, and rapid simulation and adjustment feedback \ref{design:nfr:performance}; OpenDT should enable versatile monitoring and operation, and contain multi-layer metrics for performance, sustainability, and availability~\ref{design:nfr:metrics}.
    
    \item \label{design:uc2:research} \label{design:uc2} \textbf{Researching ICT infrastructure}:
    OpenDT should allow scientific researchers to conduct accurate~\ref{design:nfr:accuracy}, rapid~\ref{design:nfr:performance}, and cost-efficient, multi-layered what-if analysis of ICT infrastructure under workload~\ref{design:nfr:metrics}. We envision future research in simulation real-time recalibration~\ref{design:fr:recalibration} and how various policies impact twinning accuracy~\ref{design:nfr:accuracy}, thus influencing practitioners in decision-making processes, and, ultimately, performance, sustainability, and availability~\ref{design:nfr:metrics}.
    
    \item \label{design:uc3:education} \label{design:uc3} \textbf{Education}: 
    OpenDT should also serve as educational material, helping future generations of capable and responsible computer scientists and engineers. OpenDT, as an open-source scientific instrument, should enable students to conduct large-scale experiments on regular-user machines, and be open-sourced, allowing students to dissect OpenDT's operational paradigm, and various metrics are twinned and predicted~\ref{design:nfr:metrics}.
    
\end{enumerate}

Based on the established use cases, we identify the following \textit{design requirements}:

\begin{enumerate}[label=\textbf{(FR\arabic*)},leftmargin=0pt,itemindent=3em]
    \item \label{design:fr:digital-twin} \textbf{Digital-twinning ICT}: \ourtool{} should ensure active and continuous replication of real-world ICT infrastructure, through a digital twin, continuously updated through telemetry data.


    \item \label{design:fr:des-simulator} \textbf{State-of-the-art, discrete-event simulation}: 
    \ourtool{} should adopt a peer-reviewed, discrete-event simulator, able to predict performance, sustainability, and availability of datacenters under workload, at a user-established granularity.

    \item \label{design:fr:recalibration} \textbf{Simulator real-time re-calibration}: 
    \ourtool{} should recalibrate predictions in real-time, based on quantified differences between predictions and real-world measurements.

\end{enumerate}

\begin{enumerate}[label=\textbf{(NFR\arabic*)},leftmargin=0pt,itemindent=3em]
    


    \item \label{design:nfr:accuracy} \textbf{Accurate, ground-truth adjusted predictions}: 
    \ourtool{} should stay within 10\% error rate (community-accepted~\cite{DBLP:conf/wosp/NiewenhuisTIM24, nicolae5377101m3sa, DBLP:conf/ccgrid/MastenbroekAJLB21, DBLP:journals/fgcs/MastenbroekMBI25}), 
    at $\ge$90\% of the operational time, with dynamic simulation re-calibration~\ref{design:fr:recalibration}.
    Inaccurate predictions can influence C-level officers to make wrong decisions~\cite{nicolae5377101m3sa}, and lead to costly downtimes~\cite{site:GoogleOutage2013}, impactful outages~\cite{site:WichitaHospitalOutage2023}, and datacenter shutdowns~\cite{site:OverheatingDataCenterOutage2023}.

    \item \label{design:nfr:performance} \textbf{Performant, lightweight digital twinning}: 
    \ourtool{} must be able to twin 7 days of real-world operation in under 1 hour, measured on a common machine (i.e., not a supercomputer). This supports live decisions for the engineering team.
    

    \item \label{design:nfr:metrics} \textbf{Multi-layer metrics}: 
    \ourtool{} must quantify metrics across performance and sustainability. For each, it must provide at least two metrics. This supports diverse scenarios.

\end{enumerate}

\begin{figure}[t]
    \centering
    \includegraphics[width=0.99\linewidth]{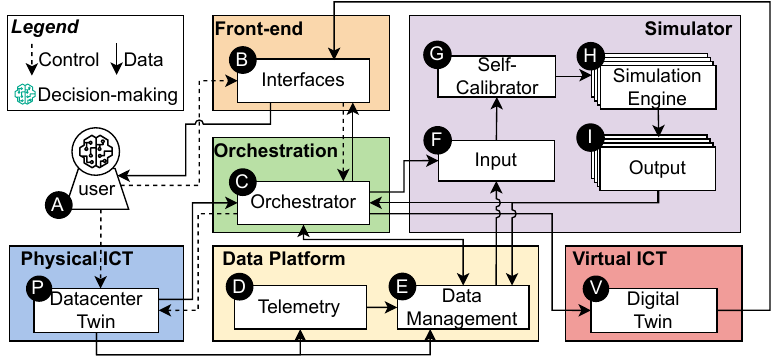}
    \vspace*{-0.35cm}
    \caption{High-level overview of \ourtool{}.}
    \Description{A high-level overview diagram of \ourtool{}.}
    \vspace*{-0.5cm}
    \label{fig:design_opendt_hl}
\end{figure}

\vspace*{-0.25cm}
\subsection{High-level design of \ourtool{}}\label{sec:design:hl-overview}

We design for \ourtool{} the architecture depicted in \Cref{fig:design_opendt_hl}. 
Overall, \ourtool{} provides a \textit{digital twin} state~(label~\circled{V}) that constantly mimics the state of the \textit{physical twin}~(\circled{P}) and is updated with what-if analysis results.
The main process supported by this architecture is for closed-loop digital-twinning~\ref{design:fr:digital-twin}, where each iteration 
(1) begins with telemetry data updates reflecting the physical datacenter's state~(\circled{P}), 
(2) then, telemetry data is fed into a discrete-event simulator~(\circled{H}) for multi-metric datacenter analysis, and into the independent self-calibration process~(\circled{G}), and
(3) the closed-loop ends with datacenter adjustments, suggested by \ourtool{}~(\circled{G}), and validated and enforced by a human-operator~(\circled{A}) for major changes. (Automating the steering process, label~\circled{D} in \Cref{fig:high-level-element}, requires careful interfacing with various kinds of physical ICT resource managers and policies, and is beyond the scope of this work.)

A key design choice for \ourtool{} is 
the simulator at its core, which is responsible for delivering high-quality predictions that represent as precisely and accurately as needed the ICT infrastructure's behaviour.
\ourtool{} uses the state-of-the-art \oursim{}~\cite{DBLP:conf/ccgrid/MastenbroekAJLB21}, a peer-reviewed, open-source, discrete-event simulator~\ref{design:fr:des-simulator}, with over 7 years of deployment and operation~\cite{DBLP:conf/ccgrid/MastenbroekAJLB21, DBLP:journals/fgcs/MastenbroekMBI25, nicolae5377101m3sa}. We augment \oursim{}'s predictive model with a real-time calibration process, based on quantified error rate between simulation and reality~\ref{design:fr:recalibration}.

To manage the challenges of the main process, where complexity is compounded by data-intensive telemetry and by compute-intensive simulation, \ourtool{} adopts an \textit{orchestrator-centric architecture}.
The Orchestrator (component~\circled{C}) operates as a central authority for task scheduling and resource management, and system health monitoring, see \Cref{sec:design:orchestrator}. 

The \ourtool{} twinning process 
is based on \textit{windows of operation}, which are short periods of time during which the system receives telemetry data (asynchronously), runs simulations, processes results, (asynchronously) updates the UI facing the human-in-the-loop, and then continues to the next time window.
Each window of operation
begins with the \textit{datacenter twin}~(\Cref{fig:design_opendt_hl}, label \circled{P}), which generates telemetry data across multiple operational layers, either directly observable metrics (e.g., instantaneous power draw, available devices), or derived information (e.g., task progress, operational phenomena), ingested by Telemetry~(component~\circled{D}). 
\ourtool{} pre-processes the ingested telemetry, converting it to simulator-ready formats and clipping data outside the window of operation, and stores it in the \textit{data management} component~(\circled{E}). 
Component~\circled{E} links various data-intensive components, and addresses the scale and frequency of telemetry data through a specialized solution---here, a Parquet-based shared file-storage was enough.

Continuing the sequence in the window of operation, \ourtool{}'s orchestrator retrieves the latest collected telemetry data and feeds it into the simulator, where the \textit{input}~(\Cref{fig:design_opendt_hl}, component \circled{F}) consists of two main data entries: 
(1) historical telemetry and past predictions, for simulation calibration~(\circled{G}), and 
(2) the latest telemetry for updating the datacenter state and for predicting future behaviour. 
One or more \textit{simulation engines}~(\circled{H}) begin making predictions, in parallel, and each produces \textit{simulation output}~(\circled{I}). 
\Cref{sec:opendt:simulator-calibrator} addresses the interplay between simulation and calibration.

The \textit{orchestrator}~(\circled{C}) publishes predictions to the \textit{front-end interface}~(\circled{B}), where the overseeing user~(\circled{A}) can intervene in decision-making processes. Ultimately, also planned work for a future \ourtool{} version, the \textit{orchestrator}~(\circled{C}) would redirect the changes to the \textit{datacenter twin}~(\circled{P}), which modifies the physical ICT based on \ourtool{}'s recommendations. 

\ourtool{} supports the human-in-the-loop with several important capabilities, including scenario configuration, automated calibration processes, and multi-model simulation techniques. We detail these capabilities and the operational workflow in \Cref{sec:design:hitl}.

\subsection{Orchestrator (Component C in \Cref{fig:design_opendt_hl})}
\label{sec:design:orchestrator}

The Orchestrator acts as a central authority and is responsible for managing core operations (e.g., simulation), system monitoring, and ensuring the correctness and coherence of the operational flow of the \ourtool{} system for each operational cycle. 

Time-wise, the Orchestrator manages the execution cycle around windows of operation, which are of fixed duration and lead to a \textit{lock-step, synchronized schedule}. This approach mitigates issues such as data misalignment and ambiguity in simulation step assignments. Without this, data, which is produced by the physical twin and does not arrive all at once in the digital twin, situations could become unclear. Similarly, repeatable trace-based simulation requires unambiguous decisions about simulation inputs. 

During each window, the Orchestrator~(\Cref{fig:design_opendt_hl}, component~\circled{C}) retrieves pre-processed telemetry from the Data Platform (via~\circled{E}) and starts the simulator with a consistent view of the datacenter state. The Orchestrator also records metadata, such as when a simulation run started and which outputs belong together, which enables correctness and performance analysis of \ourtool{} itself.

\ourtool{} supports a configurable \textit{acceleration factor}, expressed as a ratio between simulation and real-world time,
with three main modes:
(1) Simulation run one-to-one to real time (factor set to 1), 
(2) Fixed acceleration, e.g., factor set to 10, and
(3) Maximum acceleration allowed by computational resources. 
The first mode is useful for live twinning. 
The accelerated modes enable faster simulation and exploration of long-term scenarios, but require prior knowledge of the full workload trace and datacenter configuration.

In the current approach, the orchestrator does not manage \ourtool{}’s own resource allocation, nor the way its components are run. This design choice is deliberate: internal scheduling is delegated to the execution environment, allowing the orchestrator to focus exclusively on validating the digital-twinning loop itself. By keeping its functionality simple, the orchestrator ensures smooth integration among the simulation engine, data platform, and user-facing services while keeping operations interference-free.

\begin{figure}
    \centering
    \includegraphics[width=\linewidth]{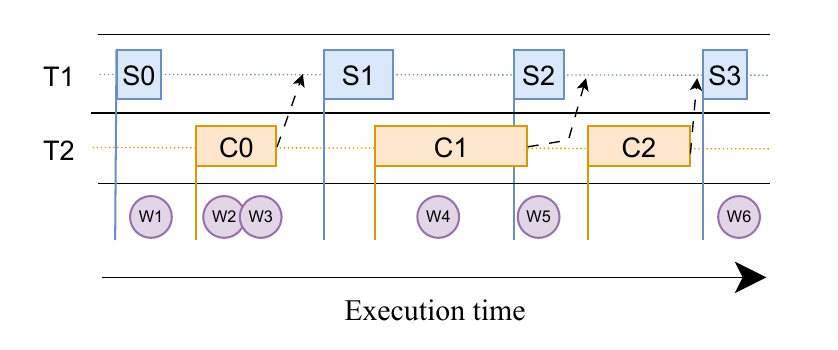}
    \vspace*{-0.75cm}
    \caption{Synchronization between simulator~(service thread~$T1$) and calibrator~($T2$). The calibrator runs~(events labeled $C$), then provides real-time feedback~(dashed arrows) to the next simulation run~($S$), enabling dynamic recalibration based on the incoming workload tasks~($W$).}
    \vspace*{-0.5cm}
    \label{fig:calibration:synchronization}
\end{figure}

\subsection{Simulator~(H) and Calibrator~(G) interplay}\label{sec:opendt:componentg}\label{sec:opendt:simulator-calibrator} \label{sec:design:simulator}

\ourtool{} uses a Simulation Engine~(\circled{H}) to inform operational decisions. Even after selecting a state-of-the-art simulator for this purpose, OpenDC~\cite{opendc-workload}, simulating large-scale and highly heterogeneous ICT infrastructure with high accuracy, precision, and explainability remains a critical yet non-trivial open scientific challenge in computer systems~\cite{DBLP:conf/ccgrid/MastenbroekAJLB21, nicolae5377101m3sa, FutureNetworkServices2025}.

Like many state-of-the-art simulators in the field, OpenDC uses a deterministic and static simulation strategy. However, static simulation models can drift and introduce errors as the time horizon increases: hardware behavior varies with temperature, aging, and firmware updates, while workload characteristics evolve over time. These dynamics threaten the validity of assumptions underlying the static model. To mitigate this problem, \ourtool{} adds a Self-Calibrator~(\circled{G}), which measures the difference between simulation-predicted results and actual telemetry, then continuously adjusts the simulation model to achieve the accuracy target~\ref{design:nfr:accuracy}.

The Self-Calibrator~(SC) employs a grid search strategy over the parameter space of the power model. For each calibration cycle, the SC evaluates a fixed set of candidate parameter values, running short simulations with each configuration and comparing results against the most recent historical telemetry data. The configuration yielding the lowest Mean Absolute Percentage Error~(MAPE) is selected and transmitted to the Simulation Engine~(SE) for use in subsequent prediction windows.

\Cref{fig:calibration:synchronization} illustrates the interplay between the SE and SC, which run as parallel processes. The calibrator completes its grid search over one time window (e.g., C0 in the figure) and sends the best-found configuration to calibrate the simulator for the next window (S1). This pipelined approach allows calibration to proceed without blocking simulation. We implement the SC as a separate service to enable independent scaling and inspection; we discuss the resulting synchronization challenges in \Cref{sec:implementation:calibration}.

In \Cref{sec:experiments:2-calibration}, we demonstrate experimentally that this calibration approach improves simulation accuracy both over uncalibrated \ourtool{} and over state-of-the-art tools~\cite{DBLP:conf/wosp/NiewenhuisTIM24,nicolae5377101m3sa}.

\begin{figure}
\centering
    \centering
    \makebox[\linewidth][c]{%
        \includegraphics[width=1.1\linewidth]{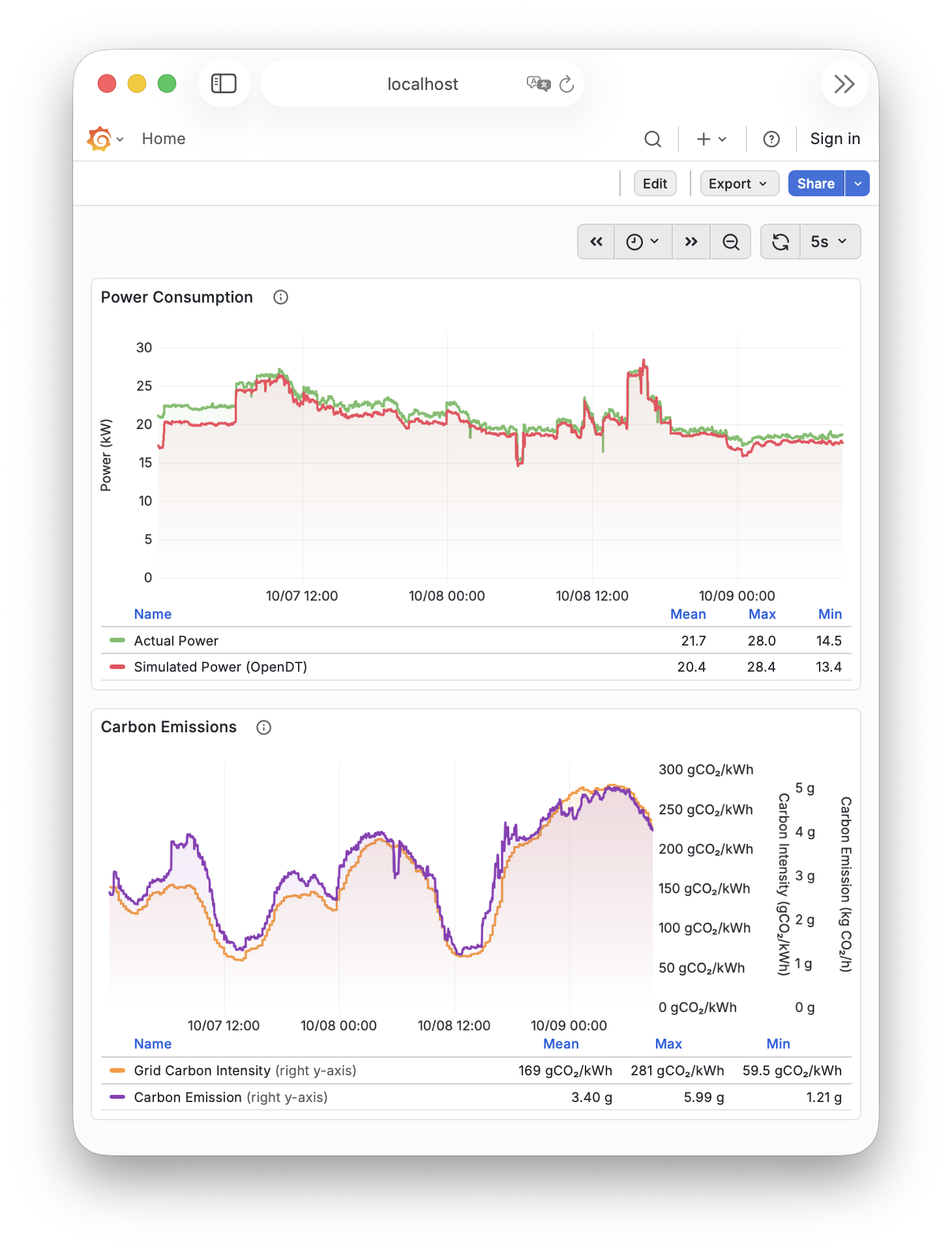}%
    }
    \vspace*{-1.1cm}
    \caption{The OpenDT Grafana dashboard.}
    \vspace*{-0.5cm}
    \label{fig:grafana-dashboard}
\end{figure}

\newcommand{\cellt}[1]{\parbox[t]{\linewidth}{#1}}

\begin{table*}[t]
  \centering
  \caption{OpenDT API endpoints for datacenter topology management and retrieving simulation metrics. The API uses JSON for all request and response bodies, with timestamps in ISO~8601 format.}
  \label{tab:api-endpoints}
  \normalsize
  \setlength{\tabcolsep}{5pt}
  \renewcommand{\arraystretch}{1.2}
  \begin{tabularx}{\textwidth}{@{}
    p{0.12\textwidth}
    p{0.22\textwidth}
    p{0.15\textwidth}
    p{0.15\textwidth}
    X
  @{}}
    \toprule
    \textbf{Endpoint} & \textbf{Description} & \textbf{Request Body} & \textbf{Response Body} & \textbf{Design Rationale} \\
    \midrule

    \texttt{PUT}\newline\texttt{/api/topology} &
    Update the simulated datacenter topology. Validates the structure and publishes it to Kafka for the simulator. &
    \texttt{Topology} object:\newline
    \texttt{clusters[]} containing\newline
    \texttt{hosts[]} with \texttt{cpu},\newline
    \texttt{memory}, \texttt{cpuPowerModel} &
    Confirmation with topology details, or validation error. &
    PUT with full topology state enables stateless updates from any interface (web UI, CLI, external systems) without requiring the server to track partial changes.
    \\
    \midrule

    \texttt{GET}\newline\texttt{/api/power} &
    Retrieve aligned power usage time series comparing simulated and actual consumption. &
    Query params:\newline
    \texttt{interval\_seconds}\newline(1--3600)\newline
    \texttt{start\_time} &
    \texttt{PowerDataResponse}:\newline
    \texttt{data[]} of \{\texttt{timestamp},\newline
    \texttt{simulated\_power},\newline
    \texttt{actual\_power}\} &
    Dedicated endpoint prevents transmitting unneeded carbon data; configurable interval supports dashboard polling.
    \\
    \midrule

    \texttt{GET}\newline\texttt{/api/co2\_emission} &
    Retrieve carbon emission time series with grid intensity and power draw. &
    Query params:\newline
    \texttt{interval\_seconds}\newline(1--3600)\newline
    \texttt{start\_time} &
    \texttt{CarbonDataResponse}:\newline
    \texttt{data[]} of \{\texttt{timestamp},\newline
    \texttt{carbon\_intensity},\newline
    \texttt{power\_draw},\newline
    \texttt{carbon\_emission}\} &
    Separate endpoint avoids payload overhead when consumers only need sustainability metrics.
    \\

    \bottomrule
  \end{tabularx}
\end{table*}

\subsection{Interfaces (Component B): API and UI}\label{sec:design:api} \label{sec:design:frontend} \label{sec:design:ui}

\ourtool{} enables both monitoring and exploratory analysis on the target system. To this end, we design a simple API interface and an interactive visual component as the front-end~(\circled{B}) to display the state of the digital twin~(\circled{V}). 

The API follows three key design principles: (1)~\emph{separation of concerns}, where simulation and data retrieval are compartmentalized and independent of each other; (2)~\emph{stability}, granted by only using a simple set of endpoints that are well-defined and documented; and (3)~\emph{observability}, achieved by showing the simulated and measured metrics in comparison. \Cref{tab:api-endpoints} summarizes the API's main endpoints for topology management and metric retrieval, including the detailed data formats for each.

\paragraph{API Design Choices.}
The topology endpoint uses the \texttt{PUT} HTTP method with a complete topology object in the request body. This design enables stateless topology updates: any client (whether a web dashboard, command-line tool, or external orchestration system) can submit a full datacenter configuration without the API server needing to maintain session state or track incremental changes. The topology schema supports hierarchical modeling with clusters containing hosts, where each host specifies CPU, memory, and power model parameters.

For metric retrieval, we provide separate \texttt{GET} endpoints for power consumption and carbon emissions rather than a single combined endpoint. This separation prevents unnecessary data transmission when a consumer requires only one metric type. For example, a real-time power monitoring dashboard can poll \texttt{/api/power} without receiving carbon intensity data it does not display. Both endpoints accept optional query parameters for sampling interval and start time, allowing clients to request data at the granularity appropriate for their use case, from fine-grained analysis (1-second intervals) to aggregated dashboard views (up to 1-hour intervals).

The API is implemented using FastAPI~\cite{fastapi} with built-in documentation rendered via Swagger~\cite{swagger}, providing interactive endpoint exploration and request testing for developers integrating with \ourtool{}.

\paragraph{User Interface.}
The main user interface is constructed with Grafana, a state-of-the-art visualization tool~\cite{grafana}. It queries the API to display simulated and measured signals side-by-side, providing a comparative view of system performance. \Cref{fig:grafana-dashboard} shows the \ourtool{} dashboard, which presents two primary panels: (1)~a power consumption panel displaying actual power draw alongside simulated predictions from \ourtool{}, enabling operators to assess simulation accuracy in real time; and (2)~a carbon emissions panel correlating grid carbon intensity with the resulting emission rates. The human-in-the-loop can inspect system state, identify periods of simulation drift, or reconfigure the profile of the simulated ontology through the topology API. As new data arrives in real-time, performance, energy consumption, and effects of calibration can be monitored continuously.

\subsection{Data Platform (Components D and E)}
\label{sec:design:dataplatform}

The \textit{Data Platform}~(\Cref{fig:design_opendt_hl}, components \circled{D}, \circled{E}) manages telemetry and digital twinning data throughout the operational cycle.

\paragraph{Telemetry Ingestion.}
The Telemetry component~(\circled{D}) continuously receives state-updates from the physical twin representing \textit{observable telemetry}, often captured through multiple lenses; this contrasts to \textit{unobservable telemetry}, which exists in reality but cannot be captured due to privacy or monitoring limitations. In \Cref{fig:design_opendt_hl}, the arrow connecting~\circled{D} to~\circled{E} represents data (pre)processing, either for further usage (e.g., simulation, calibration), or for storage and logging. The telemetry component converts raw monitoring data into simulator-ready formats and clips data outside the current window of operation.

\paragraph{Storage Format.}
The simulation results and telemetry data are persisted in Parquet format~\cite{ApacheParquet}, a columnar storage format designed for analytical workloads. We selected Parquet for four reasons: (1)~\emph{scalability}, as columnar storage enables efficient querying over large time-series datasets; (2)~\emph{storage efficiency}, through built-in compression that reduces disk footprint; (3)~\emph{cross-system compatibility}, allowing data to be consumed by diverse tools including Python (Pandas, PyArrow), Spark, and database systems; and (4)~\emph{portability}, enabling straightforward data export for external analysis or archival. The API microservices~(\Cref{sec:design:api}) query this Parquet-based storage to provide the human-in-the-loop~(\circled{A}) access to simulated power, sustainability, and other relevant metrics through HTTP endpoints, maintaining a clean separation between the simulation engine and data access layers.

\subsection{Human-in-the-Loop (Component A)} \label{sec:design:humanintheloop}\label{sec:design:hitl}

\ourtool{} adopts a human-in-the-loop~(HITL) design, where \ourtool{} operates autonomously, yet a human can (optionally) intervene and dictate major decisions. Alternatively, the system could operate fully autonomously, isolated from external input, where only SLOs dictate. However, we argue that only the former is ethically viable, especially for ICT under critical workloads, where the system could prioritize meeting SLOs over human safety or well-being (e.g., prioritizing high-paid workload from industry instead of low-paid, life-dependent workload from hospitals)~\cite{Iosup2024DigitalTwins}. Although mis-prioritization is theoretically preventable through perfect systems and SLOs, distributed systems are far from perfect, and even small operational errors can cause critical disruptions~\cite{Iosup2024DigitalTwins}.

\paragraph{HITL Capabilities.}
\ourtool{} supports three categories of human-in-the-loop interaction. First, \emph{scenario configuration}: users can define telemetry data streams, select which metrics to monitor, and specify multi-metric analysis pipelines. Configuration can occur offline through JSON topology files before the twinning process starts, or at runtime through the REST API~(\Cref{sec:design:api}), enabling dynamic adjustments without restarting the system. Second, \emph{automated calibration with human oversight}: while the self-calibrator~(\Cref{sec:opendt:simulator-calibrator}) autonomously adjusts simulation parameters, the operator retains visibility into calibration decisions through the dashboard and can override parameters via topology updates if the automated adjustments produce undesirable results. Third, \emph{multi-model simulation}: operators can configure \ourtool{} to run multiple heterogeneous power models in parallel~\cite{nicolae5377101m3sa}, comparing their predictions to identify model-specific biases and select the most appropriate model for their infrastructure characteristics.

\paragraph{Operational Workflow.}
The human-in-the-loop interacts with \ourtool{} through a continuous feedback cycle involving the Grafana dashboard and the topology API. During normal operation, the operator monitors the dashboard~(\Cref{fig:grafana-dashboard}), which displays both actual measurements from the physical datacenter and simulated predictions from \ourtool{} side-by-side. This comparative view enables the operator to assess simulation accuracy in real time: when the simulated power curve closely tracks the actual power consumption, the digital twin is accurately representing the physical infrastructure; when drift occurs, the operator can identify whether intervention is needed.

When the operator wishes to explore configuration changes or correct simulation parameters, they submit an updated topology through the API~(\Cref{sec:design:api}). The updated topology, which may include modified host counts, adjusted CPU power model parameters, or reconfigured cluster structures, is validated and published to the message queue. The simulator picks up the new topology for subsequent simulation windows, and the resulting predictions are written to the data platform. The dashboard, polling the API at regular intervals, then reflects these updated predictions. This closed-loop interaction allows the operator to iteratively refine the digital twin: submit a topology adjustment, observe its effect on predicted metrics, and adjust further if necessary.

\paragraph{Decision Support.}
Beyond monitoring, the HITL design supports proactive decision-making. For example, an operator planning a hardware upgrade can first model the change in \ourtool{} by updating the topology with the proposed new host specifications. The simulator then predicts the power consumption and carbon emissions under the current workload with the hypothetical hardware. By comparing these predictions against the baseline, the operator gains quantitative insight into the expected impact before committing resources to the physical change. Similarly, operators can use \ourtool{} to evaluate scheduling policy changes, capacity planning scenarios, or the sustainability implications of different power sourcing strategies, all while maintaining oversight and final approval authority over any changes propagated to the physical infrastructure.

\section{Implementation} \label{sec:implementation}

\begin{figure}[t]
    \centering
    \includegraphics[width=\linewidth]{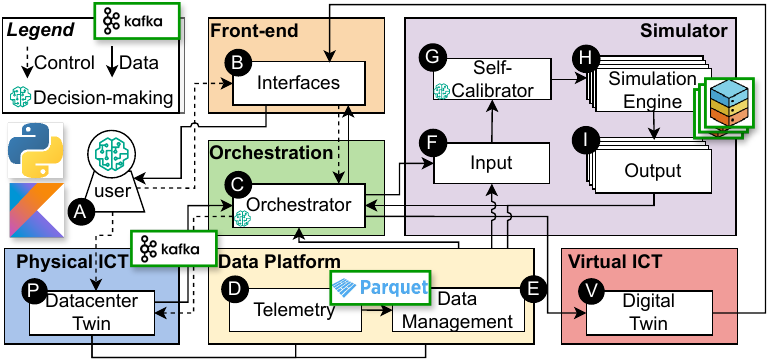}
    \caption{A high-level overview of the software stack in the current OpenDT engineered prototype.}
    \label{fig:design:technologies}
\end{figure}

Following the system design, we will now describe the implementation of OpenDT, which provides the concrete system on which the experiments in~\Cref{sec:experiments} are carried out. 
Instead of documenting every component individually, we present the main challenges and solutions encountered in the engineering process.

\Cref{fig:design:technologies} illustrates \ourtool{}'s software stack. Overall, OpenDT is written in Python and is deeply embedded with \oursim{}, written in Kotlin.
The implementation is built as a Docker Compose microservice system in which the core components (data source, simulator, calibrator, and API) communicate through Apache Kafka~\cite{kafkapaper}.
All services mount a shared host directory that serves as a file-based workspace for configuration files and simulation outputs. This shared location makes runtime data easy to inspect, and helps with reasoning about the persisted state of the system.

\subsection{Retrofitting OpenDC for Online Digital Twinning}
\label{sec:implementation:opendc}

The OpenDC simulator is originally designed as an offline simulator that processes an entire workload 
in a single execution. In OpenDT, however, workload data arrives continuously, and the simulator must generate updated predictions at fixed intervals. Supporting this online, incremental setting required a redesign of how workloads are aggregated and supplied to OpenDC.

OpenDC models each task as a quantity of compute cycles, which is then subdivided into fragments that describe its expected CPU usage over time. 
In an online setting, tasks may span multiple prediction windows or originate much earlier while still influencing current CPU utilization and power behaviour. 
For this reason, OpenDT invokes OpenDC on an accumulated set of all tasks received up to the current simulated time, ensuring that any task that could potentially affect the present window is included.

The drawback of this approach is a growing context horizon, since each OpenDC invocation receives all previously observed tasks. 
A more efficient implementation would incorporate task completion information from earlier simulation outputs so that future invocations can exclude workloads that have already finished. This remains future work.

\subsection{High Throughput Workload Ingestion via Kafka}\label{sec:implementation:kafka}
To support fast experimentation~\ref{design:nfr:performance}, we added a feature to replay the workload traces at a configurable speed factor. This pragmatic choice reduced iteration time but also surfaced limitations in the initial ingestion design. In the first version, tasks and their corresponding fragments were published as separate Kafka messages on two topics. Although this mirrored the structure of the raw trace dataset, it required two producers to stay aligned with the simulated timeline. Since workload traces contain far more fragments than tasks, Kafka struggled to keep up at higher replay speeds. Fragment messages accumulated, arrived late at consumers, and required additional logic to associate them with previously received tasks.

To eliminate this coordination problem, we redesigned ingestion around a single nested task object. The workload generator performs a one-time join between each task and its fragments and sends a single Kafka message containing the complete structure. This removes the fragment topic entirely and reduces the message volume to the number of tasks. It also allows ingestion to proceed through a single producer thread, which avoids the timing issues present in the original design and produces a more predictable ingestion pipeline.

\subsection{Progress Markers: Deliverability \& Window Completeness} \label{sec:implementation:heartbeat}
The simulator processes workload traces at a fixed simulation time frequency. Implementing this reliably required a mechanism to determine when all submissions belonging to a time window had arrived. Without such a guarantee, the simulator might either miss late-arriving tasks or inconsistently adjust window boundaries, which either impacts reproducibility or complicates downstream analysis.

To address this, the workload producer sends periodic progress markers in addition to task submissions. These markers appear at fixed simulated intervals, for example, once per minute. Since Kafka preserves message order within a partition, the arrival of a marker for timestamp $t$ indicates that all submissions with earlier timestamps have been sent. The simulator uses these markers to trigger a simulation step only after the corresponding window is complete. This approach removes the dependence on runtime timing and ensures consistent formation of simulation windows.

\subsection{Calibration and Temporal Synchronization}\label{sec:implementation:calibration}
A second major design decision involved the calibration mechanism. We considered integrating the calibrator directly into the simulator, which would have simplified data sharing and ensured a single ordered execution flow. However, this would have tightly coupled the components, prevented independent resource scaling, and made inspection of each component more difficult.

For this reason, we implemented the calibrator as a separate service that consumes the same workload topics and publishes proposed topology updates. This decoupling introduced the challenge of temporal alignment. To address this, we implemented a best-effort synchronization mechanism. Each service tracks its own perceived simulated time; if it runs ahead, it delays execution until the next window boundary, and if it falls behind, it logs warnings because there is no automatic way to accelerate processing. When this happens, the system operator is expected to adjust the speed factor accordingly.

The decoupled and asynchronous behaviour is illustrated in \Cref{fig:calibration:synchronization}. The figure shows how the simulator and calibrator process the workload stream independently, each advancing through its own sequence of workload tasks, with the calibrator sending updated topology information back to the simulator.

Although approximate, this approach keeps the simulator and calibrator sufficiently aligned to produce consistent calibration results and stable system behaviour in practice. With these mechanisms in place, the prototype offers a reproducible execution environment that supports the trace-based experiments presented in the following section.

\begin{table*}[t]
\small
\centering
\caption{Experiment configurations. SUO = system under observation, WT = workload trace, PM = performance metrics, SM = sustainability metrics, OP = operational phenomena (failures), RSR = real-time simulation recalibration. Time = workload execution time, CPU = average CPU utilization [\%], }
\label{table:experiments-overall-design}
\begin{tblr}{
  column{1,2,9-11} = {l,m},     
  column{3-8}      = {l,m},     
  column{6} = {rightsep=6pt},
  column{7} = {leftsep=6pt},
  column{9-11}     = {c},
  cell{1}{4} = {c=3}{c},        
  cell{1}{7} = {c=2}{c},        
  row{2} = {},                  
  hline{1,3,8} = {-}{},
  hline{2} = {4-8}{}
}
        &                                                                                                   &    & Input   &     &     & Output &     &     &     &     \\
Section & Focus                                                                                             &    & SUO     & WT         &  & PM     & SM  & OP  & RSR \\
$\S$\ref{sec:experiments:exp-1} & {Reproduce peer-reviewed\\experiment with twinning capabilities}             &    & S1      & WT1        & & {time, CPU} & {Wh, gCO2}  & \ding{56} & \ding{56} \\
$\S$\ref{sec:experiments:exp-2} & {Evaluate simulation calibration \\ against current state-of-the-art}        &    & S1      & WT1        & & {\ding{56}} & {Wh}        & \ding{56} & \ding{52}  \\
\hline
\end{tblr}
\end{table*}

\begin{table}[t]
\small
\centering
\caption{Workload traces used in experiments. Name is source and collection year (e.g., SURF-22 = source SURF, year 2022). SR = sampling rate, CH = CPU hours (millions), S = scientific.}
\label{table:workload-traces}
\begin{adjustbox}{width=\columnwidth,center}
\begin{tblr}{
  column{5} = {r},
  column{6} = {r},
  column{7} = {r},
  column{8} = {r},
  cell{1}{5} = {c},
  cell{1}{6} = {c},
  cell{1}{7} = {c},
  cell{1}{8} = {c},
  hline{1-2,6} = {-}{},
}
ID  & Name         & Type & Duration & Jobs  & CH        & SR  \\
WT1 & SURF-22      & S    & 7\,days    & 7,850 & 0.31   & 30\,s \\
\hline
\end{tblr}
\end{adjustbox}
\end{table}
\begin{table}[t]
\small
\centering
\caption{Systems Under Observation (SUO). Scale = number of hosts.}
\label{table:systems-under-observation:suos}
\begin{tblr}{
  column{3} = {r},
  hline{1-2,5} = {-}{},
}
ID & Source  & Scale & Resources per host            \\
S1 & SURF    & 277             & 128~GB RAM, 16 cores, 2.1~Ghz \\
\hline
\end{tblr}
\end{table}

\section{Trace-based experiments, enabled by and conducted with \ourtool{}} \label{sec:experiments}

In this section, we validate \ourtool{} using real-world traces, compare with a peer-reviewed simulator, and explore if, and by how much, real-time simulation recalibration improves accuracy. Overall, our experiments support \textit{main findings~(\textbf{MF1-3})}:

\begin{enumerate}[label=(\textbf{MF\arabic*}),leftmargin=0pt,itemindent=3em]
    \item \label{experiments:mf1} 
    \ourtool{} allows for continuous replay of real-world datacenter operation, with high accuracy. Compared to the measured MAPE error rate of the peer-reviewed simulator \footprinter{} 7.86\%~\cite{DBLP:conf/wosp/NiewenhuisTIM24}, for this experiment \ourtool{} achieves a MAPE of 5.13\% ($\S$\ref{sec:experiments:exp-1}). 
    
    \item \label{experiments:mf2}
    \ourtool{}'s live self-calibration is enabled by digital twinning,  happens regularly and continuously, and 
    improves MAPE; here, from 5.13\% to 4.39\% ($\S$\ref{sec:experiments:exp-2}).
    
    \item \label{experiments:mf3}
    \ourtool{} can twin multiple core metrics across operational layers, related to sustainability (e.g., power draw, efficiency, $\S$\ref{sec:experiments:1-footprinter}) and performance (e.g., TFLOPs, $\S$\ref{sec:experiments:exp-1}, and CPU utilization, $\S$\ref{sec:experiments:2-calibration}). 

    \item \label{experiments:mf4}
    Supporting C-level decisions and datacenter operators, OpenDT can show underutilised compute infrastructure, yet consuming large amounts of electricity while being idle. One such example is illustrated in \Cref{sec:experiments:exp-1} and \Cref{fig:exp1_plot_cpu_latency}, where, by reproducing a real-world scenario from SURF-22 in silico, with OpenDT capabilities, we show that only 25\% of the available and powered-on compute power is utilised.
\end{enumerate}

\subsection{Prototype implementation and performance} \label{sec:implementation}

We implement \ourtool{} as a Docker Compose microservice system in which the core components (data source, simulator, calibrator, and API) communicate through Apache Kafka~\cite{kafkapaper}.

All services mount a shared host directory that serves as a file-based workspace for configuration files and simulation outputs. This shared location makes runtime data easy to inspect and helps with reasoning about the persisted state of the system.

For implementation details, see \Cref{sec:implementation}.

\textit{Performance}: Enabled by the lightweight nature of the OpenDT prototype, we run both experiments on a common off-the-shelf MacBook Pro, with an M1~Max~10-core CPU and 32~GB of RAM. This showcases OpenDT's capabilities of twinning 7 days of datacenter operation within 46 minutes, thus successfully addressing~\ref{design:nfr:performance}.

\subsection{Experiment setup}\label{sec:experiments:setup}

\textit{Infrastructure:} In these experiments, we twin the SURF-SARA production cluster at SURF, the Dutch infrastructure for scientific computing. SURF-SARA contains 277 hosts, each with 128~GB of RAM and 16 processing cores running at maximum 2.1~GHz. 

\textit{Workload trace (public):} We use SURF-22, a scientific workload trace from SURF, also used in peer-reviewed articled~\cite{DBLP:journals/fgcs/VersluisCGLPCUI23,DBLP:conf/wosp/NiewenhuisTIM24}.It traces scientific jobs with an average duration of 39.52\,CPU-hours~\cite{nicolae5377101m3sa}.

\textit{Quantifying accuracy / error rate:} We employ Mean Percentage Average Error (MAPE), a widely used and relative-error metric \cite{oracle2014mape, DBLP:conf/wosp/NiewenhuisTIM24, moreno2013using-mape, nicolae5377101m3sa}. MAPE is calculated as $\text{MAPE}~[\%] = \frac{1}{n} \sum_{i=0}^{n} \left| \frac{R_{i} - S_{i}}{R_{i}} \right| \times 100$, where $n$ is the number of samples, $R$ is the real-world data~\cite{DBLP:journals/fgcs/VersluisCGLPCUI23}, $S$ is the simulation data, and $i$ is the sample index.

\textit{Power model:} We model CPU power draw adopting the \oursim{}~\cite{DBLP:conf/ccgrid/MastenbroekAJLB21, DBLP:conf/isca/FanWB07} analytical formula: $P(u)=P_{\text{idle}}+(P_{\text{max}}-P_{\text{idle}})\,(2u-u^{r})$. Here, $u$ is CPU utilization, $P_{\text{idle}}$ and $P_{\text{max}}$ represent idle and max power, and $r$ is the \textit{calibration parameter}~(see~\Cref{sec:opendt:simulator-calibrator}).

\begin{figure}[t]
    \centering
    
    \includegraphics[width=\linewidth]{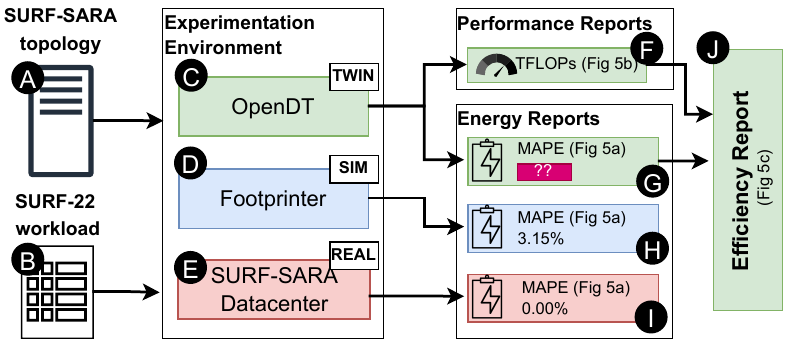}
    \caption{\Cref{sec:experiments:1-footprinter} experiment, adapted and re-run from \footprinter{}~\cite{DBLP:conf/wosp/NiewenhuisTIM24} and extended with \ourtool{}.}
    \label{fig:exp1-overview}
\end{figure}

\begin{figure}[t]
    \centering
    \includegraphics[width=\linewidth]{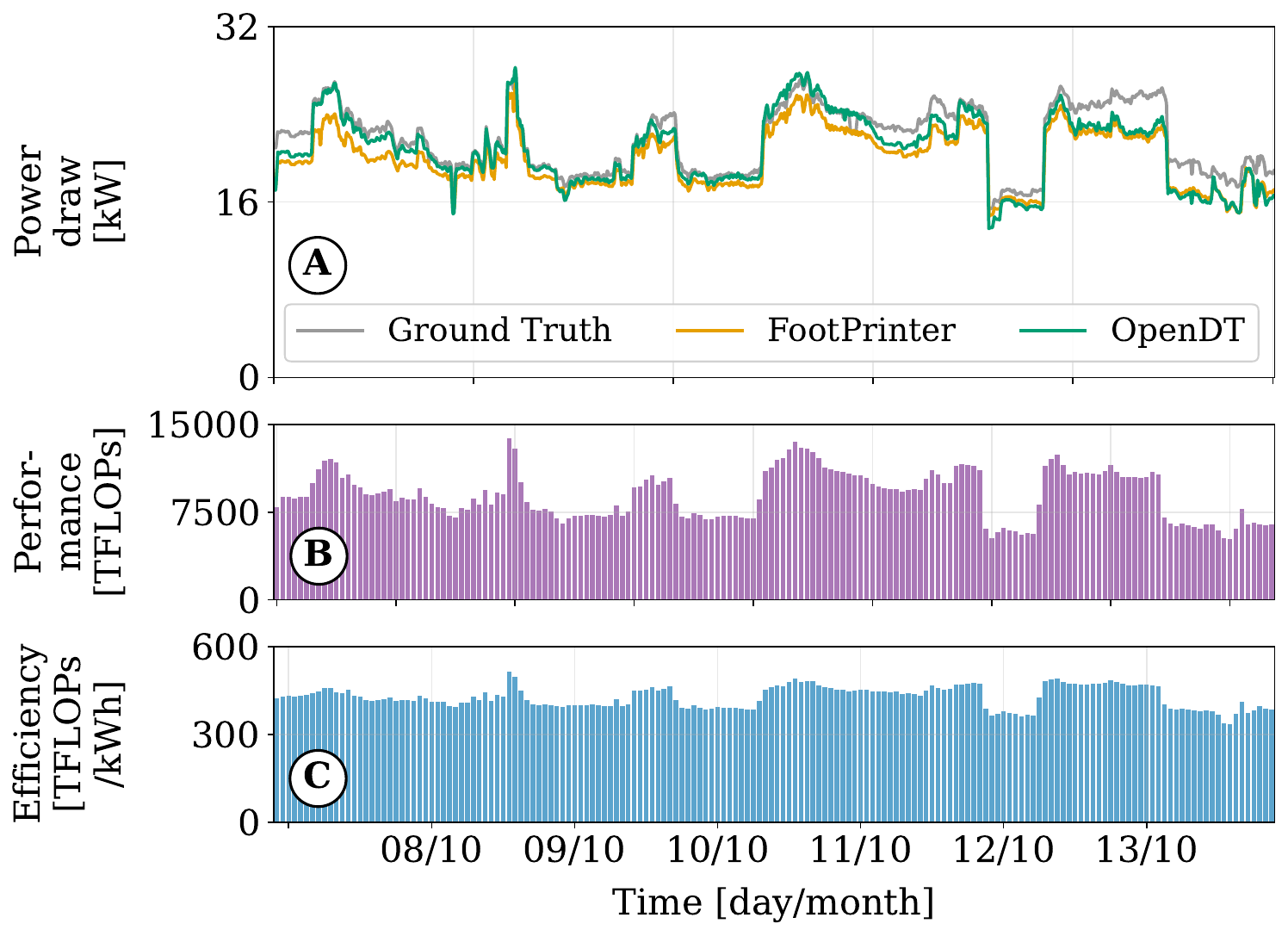}
    \caption{Operational metrics for the compute cluster over time: (\textbf{A}) Power draw simulation results vs. measured reality (reproduced from~\cite{nicolae5377101m3sa, DBLP:conf/wosp/NiewenhuisTIM24}). (\textbf{B}) Performance measured in TFLOPs. (\textbf{C}) Efficiency measured in TFLOPs/kWh.}
    \label{fig:exp1-results}
\end{figure}

\subsection{Experiment E1: Peer-reviewed experiment reproduced and extended with \ourtool{}}\label{sec:experiments:exp-1} \label{sec:experiments:1-validation} \label{sec:experiments:1-footprinter}

Niewenhuis et al. propose \footprinter{}~\cite{DBLP:conf/wosp/NiewenhuisTIM24}, a tool for predicting the CO$_2$ footprint of datacentres.
\Cref{fig:exp1-overview} illustrates the workflow we followed in this work, leveraging \ourtool{} capabilities to reproduce the key experiment~\cite{DBLP:conf/wosp/NiewenhuisTIM24}. 
Illustrative for the difference between approaches, whereas \footprinter{} (\circled{D}) runs once a hand-tuned energy model~\cite{DBLP:conf/wosp/NiewenhuisTIM24, nicolae5377101m3sa}, \ourtool{} (\circled{C}) continuously predicts energy consumption at the industry-standard sampling granularity (i.e., 5-minute rate), using a generic predictive model that avoids overfitting for a specific trace. 
To reproduce the experiment, we use datacenter topology~(\circled{A}) and workload trace~(\circled{B}) recorded
from SURF~\cite{nicolae5377101m3sa, DBLP:conf/wosp/NiewenhuisTIM24}~(\circled{E}), which we regard as ``\textit{ground-truth}''~(\circled{I}). 
We measure the MAPE of both \ourtool{}~(\circled{G}), compare it with \footprinter's~(\circled{H}), and also record datacenter performance data~(\circled{I}) that enables us to extend the experiment and also report energy-efficiency results~(\circled{J}).
The results of this experiment firmly support main findings MF1 and MF3.

\textit{Accuracy validation by reproducing the peer-reviewed experiment}: We determine the accuracy of predictions by determining the MAPE error rate, using the formula described in~\Cref{sec:experiments:setup}; the lower the MAPE, the more accurate the prediction. \Cref{fig:exp1-results}A depicts the results of the reproducibility experiment. 
Between
predictions and ground truth (measured reality, we compute \footprinter{}'s MAPE as 7.86\% and \ourtool{}'s MAPE as 5.13\%~(2.73\% better). Here, even without simulation recalibration, \ourtool{} meets the accuracy requirement~\ref{design:nfr:accuracy}.

\textit{Extending the peer-reviewed experiment}: Beyond quantifying datacenter sustainability, \ourtool{} can also predict performance and efficiency quickly. \Cref{fig:exp1-results} depicts the results obtained when we extend the \footprinter{} experiment with performance results and demonstrates that \ourtool{} meets~\ref{design:nfr:metrics}. 

In~\Cref{fig:exp1-results}B, we illustrate live, continuous predictions of \ourtool{} on datacenter performance, which are further processed to produce the efficiency evaluation depicted in~\Cref{fig:exp1-results}C. Overall, discretising \ourtool{} predictions per hour, we identify the highest efficiency when the datacenter performance, quantified in floating-point operations, is the highest. Further investigating \ourtool{}'s predictions, we identify underutilization of the available infrastructure: the monitoring period, under 30\% of the available processing power is used, while the remaining are idle. Such insights, enabled by \ourtool{}, could help operators better monitor and plan available infrastructure.

\begin{figure}[t]
    \centering
    \includegraphics[width=0.9\linewidth]{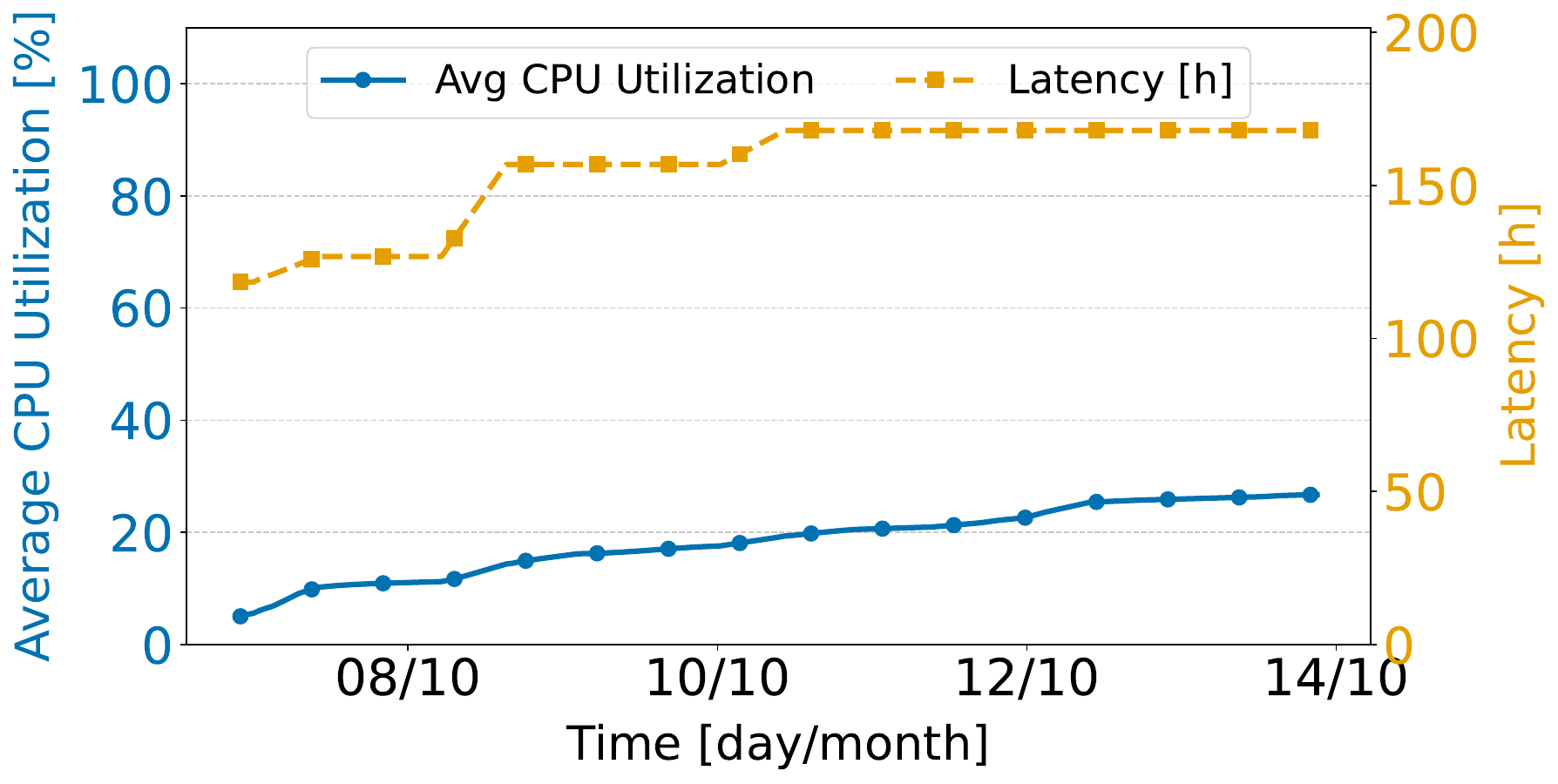}
    \caption{Performance metrics of S1 under SURF-22 workload, over time. Through digital twinning, we capture how variations in the workload can impact average CPU utilization and total runtime (i.e., how much time it takes for the datacenter to complete the workload, assuming no changes).}
    \label{fig:exp1_plot_cpu_latency}
\end{figure}

\Cref{fig:exp1_plot_cpu_latency} showcases OpenDT's capabilities of predicting two performance metrics \ref{design:nfr:metrics}, in real-time. We digital twin, with OpenDT capabilities, S1 datacenter under SURF-22 workload, and trace the average CPU utilization and latency (i.e., the estimated remaining time, until the workload is complete, assuming no change in the system state). 

\textit{Simulating latency:} We observe a slight increase in latency over time, from $\approx$60 hours to $\approx$90 hours until the workload would be finalized. OpenDT's continuous twinning capabilities showcases two main timestamps when the latency increases, potentially due to additional tasks the datacenter needs to execute. Such analysis and OpenDT-aided historical replay can help practitioners~\ref{design:uc1:practitioner} take informed decisions on datacenter management (e.g., up/down-scaling), or could pinpoint to scientists~\ref{design:uc2:research} hot operational intervals, useful for exploration.

\textit{CPU utilization:} \Cref{fig:exp1_plot_cpu_latency} emphasises a critical, yet non-trivial challenge in ICT keeping resources, in this case CPUs, idle~\cite{DBLP:conf/ccgrid/MastenbroekAJLB21, DBLP:journals/tpds/AndreadisMBI22, DBLP:journals/corr/abs-2206-03259}; an idle CPU can consume up to 40\% of the energy it would consume at maximum utilization~\cite{DBLP:conf/ccgrid/MastenbroekAJLB21, DBLP:journals/tpds/AndreadisMBI22}. Through OpenDT, we identify, in \Cref{fig:exp1_plot_cpu_latency}, an under-utilization of the available infrastructure, where only about 25\% of the available processing power is used, while the remaining three quarters are consuming power for being idle. Such insights, enabled by OpenDT, could aid operators in better monitoring and planning available infrastructure~\ref{design:uc1:practitioner}.
 
\subsection{E2: Evaluating live, self-recalibration against the current simulation state-of-the-art}\label{sec:experiments:exp-2} \label{sec:experiments:2-calibration}

\begin{figure} [t]
    \centering
    \includegraphics[width=0.98\linewidth]{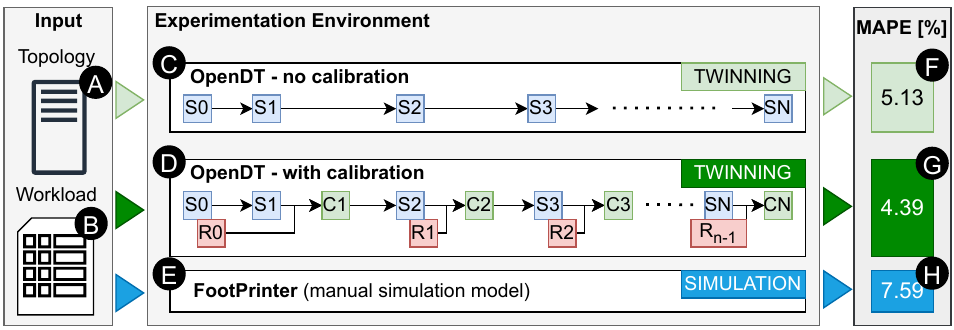}
    \caption{Overview of experiment 2. Evaluating how well OpenDT predicts, with and without real-time simulation re-calibration, in comparison with the peer-reviewed FootPrinter~\cite{DBLP:conf/wosp/NiewenhuisTIM24}. S = simulation, C = calibration, R = real-world measurements (required for dynamically, on-the-fly, recalibrating predictions).}
    \label{fig:exp2:overview}
\end{figure}

Addressing the challenge of improving prediction accuracy for digital twins~(see~\Cref{sec:design:simulator}), we evaluate the Self-Calibrator.
Experimentally, we analyze how real-time recalibration affects \ourtool{}'s accuracy relative to traditional simulation, which is not live-calibrated.

\begin{figure}[!t]
    \centering
    \includegraphics[width=\linewidth]{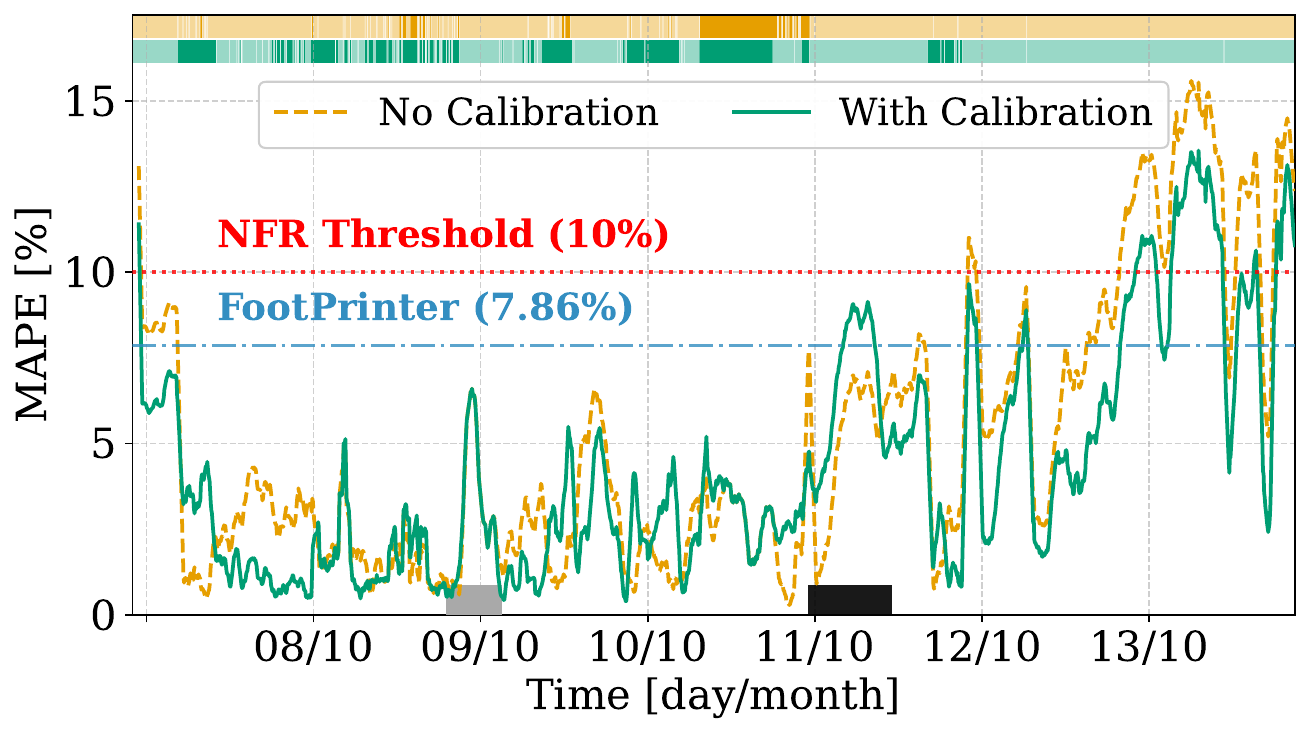}
    \caption{Evolution of error rate in power-draw estimation of \ourtool{} with calibration (MAPE 4.39\%), without calibration (MAPE 5.13\%), compared with NFR1 and \footprinter{}. The horizontal bars at the top distinguish OpenDT over-estimations (solid color) from under (dimmed), without calibration (orange, top bar) and with calibration (green). The black and grey marks at the bottom mark special intervals, see text.}
    \label{fig:exp2:mape-over-time}
\end{figure}

\Cref{fig:exp2:mape-over-time} depicts the MAPE over time for \ourtool{} with and without calibration, against the set threshold, i.e., below 10\%, 90\% of the time~\ref{design:nfr:accuracy}. Overall, the live-recalibration approach reduces MAPE by 0.74, from 5.13\% to 4.39\%. We identify for the uncalibrated \ourtool{} 
the MAPE is $<10\%$ occurs only 86\% of the time. However, the live re-calibrated \ourtool{} achieves MAPE $<10\%$ just over 92\% of the time, thus successfully meeting \ref{design:nfr:accuracy}.

\textit{Analysis of calibration vs. no calibration}: 
Although the calibration technique proposed in this work is relatively simple, the calibrated error rate of \ourtool{} is better than that of the traditional approach. It would be tempting to conclude that using calibration is always beneficial. 
However, 
we set out to investigate whether this is the case, with results depicted by \Cref{fig:exp2:mape-over-time}.
We observe that there exist significant simulation-periods when the no-calibration technique performs better than the \ourtool{} calibration (e.g., a window of about 12 hours centered in 11/10, highlighted in black in~\Cref{fig:exp2:mape-over-time}) or equally good (e.g., around 9/10, highlighted in grey). This suggests calibration is important but requires further attention.

\textit{Prediction behaviour:} 
A large body of work, including SPEC RG Cloud's work on auto-scaling metrics~\cite{DBLP:journals/tompecs/HerbstBKOEKEKBA18}, addresses the impact of under- and over-estimation in infrastructure provisioning.
Underestimating and under-provisioning ICT infrastructure could lead to performance concerns, system faults, and, ultimately, system failures. In contrast, overprovisioning could lead to sustainability concerns, both environmental and financial, where energy (and thus, money) is wasted by idle infrastructure. In~\Cref{fig:exp2:mape-over-time}, marked by the horizontal bars at the top of the figure, we identify an underestimation trend in \ourtool{}'s predictive model. This occurs both without calibration (in 85\% of discrete-event predictions) and with calibration (66\%). Overall, \textit{we find calibration alleviates underestimation bias, reducing the frequency of underestimated predictions}.

\section{Discussion} \label{sec:discussion}

In this section, we evaluate OpenDT towards the established functional and non-functional requirements~(\Cref{sec:discussion:reqs}. Then, in \Cref{sec:discussion:limitations} we present limitations of our experiments, simulation, and real-world tracing (telemetry).

\subsection{Requirement analysis}\label{sec:discussion:reqs}

In \Cref{sec:design}, we establish a set of \textit{requirements}, which guided the design and engineering of OpenDT. We now evaluate if, and how well, OpenDT matches the established functional (FR) and non-functional requirements (NFR).

\begin{enumerate}[label=\textbf{(FR\arabic*)},leftmargin=0pt,itemindent=3em]
    \item \label{discussion:fr:digital-twin} \textbf{Digital-twinning ICT}: 
    \ourtool{} ensures active and continuous replication of real-world ICT infrastructure through a \textit{two-stage} digital twin, which \textit{(1) ingests datacenter telemetry} at a user-established granularity, then \textit{(2) simulates} how the infrastructure would operate in the future (e.g., sustainability, performance), assuming no state changes. We plan future work towards a closed-loop digital twinning process, which would add a third stage where, (3) OpenDT would adjust a real-world datacenter, automatically and aligned with established SLOs, yet supervised by a human-in-the-loop, based on simulations from step (2).

    \item \label{discussion:fr:des-simulator} \textbf{State-of-the-art, discrete-event simulation}: 
    \ourtool{} uses \oursim{}, a peer-reviewed, discrete-event simulator, with over 7 years of deployment and operation, employed in scientific publications~\cite{DBLP:conf/ccgrid/MastenbroekAJLB21, DBLP:journals/fgcs/MastenbroekMBI25, nicolae5377101m3sa}, and in national-scale projects (e.g., the Dutch 6G Future Network Services~\cite{FutureNetworkServices2025}). \oursim{} is able to predict performance, sustainability, and availability of datacenters under workload, at a user-established granularity.

    \item \label{discussion:fr:recalibration} \textbf{Simulator real-time re-calibration}: 
    \ourtool{} support real-time simulation re-calibration, as described in~\Cref{sec:design:simulator}, where we detail the interplay between the simulator and the live-recalibration. We quantify in~\Cref{sec:experiments:2-calibration} the accuracy of the live-calibrated OpenDT simulation, as compared with the traditional, state-of-the-art, non-calibrated simulation from FootPrinter~\cite{DBLP:conf/wosp/NiewenhuisTIM24}; overall, calibration helps reduce MAPE from 7.86\% to 5.13\% (thus, 2.73\% better).

\end{enumerate}

Through experiments conducted in~\Cref{sec:experiments}, we explore real-world scenarios and measure OpenDT against the non-functional requirements established in~\Cref{sec:design}.

\begin{enumerate}[label=\textbf{(NFR\arabic*)},leftmargin=0pt,itemindent=3em]

    \item \label{discussion:nfr:accuracy} \textbf{Accurate, ground-truth adjusted predictions}: 
    As shown in~\Cref{sec:experiments:1-footprinter}, and supported by \ref{experiments:mf1}, \ourtool{} allows for continuous replay of real-world datacenter operation, with high accuracy. Compared to the measured MAPE error rate of the peer-reviewed simulator \footprinter{} 7.86\%~\cite{DBLP:conf/wosp/NiewenhuisTIM24}, for this experiment \ourtool{} achieves a MAPE of 5.13\% ($\S$\ref{sec:experiments:exp-1}). We envision future work in evaluating how various live-calibration techniques could improve simulation accuracy, aiming for improving the already low error rate even further.

    \item \label{discussion:nfr:performance} \textbf{Performant, lightweight digital twinning}: 
    We validate \ourtool{} against \ref{design:nfr:performance} through the nature of the experimental setup, detailed in~\Cref{sec:experiments:setup}. Overall, we conduct all experiments on a common off-the-shelf MacBook Pro, with an M1~Max~10-core CPU and 32~GB of RAM. Further showcasing performance capabilities, we measured the runtime of experiment E1 (\Cref{sec:experiments:1-footprinter}), where we twin 7 days of datacenter operation, to 46 minutes.
    
    \item \label{discussion:nfr:metrics} \textbf{Multi-layer metrics}: 
    As presented in~\Cref{sec:experiments}, and supported by~\ref{experiments:mf4}, \ourtool{} can twin multiple core metrics across operational layers, related to sustainability (e.g., power draw, efficiency, $\S$\ref{sec:experiments:1-footprinter}) and performance (e.g., TFLOPs, $\S$\ref{sec:experiments:exp-1}, and CPU utilization, $\S$\ref{sec:experiments:2-calibration}). 

\end{enumerate}

\subsection{Limitations}\label{sec:discussion:limitations}

We now analyze limitations of the experiments conducted in~\Cref{sec:experiments}, limitations of OpenDT, and limitations of digital twinning, overall, as a technique for monitoring and operating datacenters.

Evaluating potential threats to \textit{experimental validity}, we identify \textit{generality} and \textit{isolation}. Addressing \textit{(i) the generality aspect}, we acknowledge that the experiments conducted in this work analyze a single workload trace, specifically SURF-22, traced in the Dutch Supercomputer SURF, and reproduce a single peer-reviewed experiment. Thus, the system may or may not generalise to other datacenter topologies or workload types (e.g., business-specific), thus hindering external validity. Addressing the \textit{isolation aspect}, the experiments conducted in this work used simulation or twinning methods which assumed perfect isolation from operational phenomena, such as hardware failures; however, we argue that every real-world system, including the SURF-SARA cluster, is prone to operational phenomena, which we do not capture in this work. We plan to conduct further experimental work of OpenDT, and futher expand our set of experiments with more real-world traces, of various kinds (e.g., business-critical vs. scientific-specific), and with infrastructure facing operational phenomena -- all features already supported by OpenDC~\cite{DBLP:conf/ccgrid/MastenbroekAJLB21}, or by related tools using OpenDC~\cite{nicolae5377101m3sa, DBLP:journals/fgcs/MastenbroekMBI25, DBLP:journals/tpds/AndreadisMBI22}.

Evaluating potential threats to method validity, we identify several limitations where digital twinning could potentially be inferior to traditional simulation. At the core of any digital twin, including OpenDT, is the simulator; thus, the validity of the twin's predictions are directly dependent on the validity of the simulator's predictions. For OpenDT, the validity of OpenDT's predictions could, thus, be hindered by a potential threat to the correctness of the simulator's predictions. To address this limitation, we select for OpenDT the peer-reviewed and community-vetted OpenDC, currently trusted by national-wide infrastructure projects~\cite{FutureNetworkServices2025} and by European-scale projects~\cite{GraphMassivizer2025}. Still, we design and engineer OpenDT as modular, making it simple to integrate different discrete-event simulators.

\section{Related work: digital twinning and simulation} \label{sec:related}

We now contrast \ourtool{} with prior work in digital twinning across fields of science and in datacenter simulation. 

\textit{Digital twinning:} Digital twins are already widely used in large-scale sciences such as aviation and space exploration~\cite{Aydemir2020AircraftDT}, and enable coarse-grained, dynamic adjustments for the physical twin from a distance, without requiring physical access (e.g., Apollo 13 controlling from the Earth~\cite{allen2021digital}). 
Similarly, small-scale sciences (e.g., biomedical applications) use digital twins 
to enable what-if analysis and patient monitoring and treatment, where physical access is not viable (e.g., monitoring ventricular activity) using a closed-loop with a doctor-in-the-loop~\cite{sel2024building}. In contrast, for the medium-scale science of Computer Systems, 
where both detailed models exist for some objects and others are treated statistically, 
digital twins are only starting to emerge,
mainly due to the inherent intellectual and computational complexity of combining high-level abstractions with detailed system monitoring and massive amounts of telemetry; 
\ourtool{} is to-date the first such digital twin to publicly detail its design and become open source.

\textit{Simulation:} In this work, we leverage at the core of the digital twin the peer-reviewed \oursim{}, a discrete-event simulator able to predict datacenter performance, sustainability, and availability, in time and cost-efficent way~\cite{DBLP:conf/ccgrid/MastenbroekAJLB21, DBLP:journals/fgcs/MastenbroekMBI25}. \oursim{} is able to simulate with high explainability, through multiple aligned simulation models~\cite{nicolae5377101m3sa}, and with high robustness, through Meta-Models, which predict by combining predictions of other models, thus alleviating individual model biases~\cite{nicolae5377101m3sa}. 
Other seminal datacenter simulators, such as DCSim~\cite{DBLP:conf/green/GuptaGBAMV11}, CloudSim~\cite{DBLP:journals/spe/CalheirosRBRB11, DBLP:journals/spe/HewageIRB24, DBLP:conf/im/FilhoOMIF17}, or SimGrid~\cite{DBLP:conf/ccgrid/Casanova01}.
In contrast, \ourtool{} enables a new mode of operation, twinning live and closing the simulation loop, for which it proposes a novel design, implementation, and evaluation.

\section{Open and Reproducible science} \label{sec:open-science}

We release the engineered OpenDT prototype as FOSS, together with a FAIR dataset of traces, via GitHub\footnote{ \url{https://github.com/atlarge-research/opendt/tree/hcp}}, accessible by all groups of stakeholders, thus aiding \ref{design:uc1:practitioner}, \ref{design:uc2:research}, and \ref{design:uc3:education}. Adhering to state-of-the-art principles on reproducible science, the OpenDT artifact contains a reproducibility capsule.

Furthermore, addressing the main stakeholders, we ensure robust and clear documentation for OpenDT, through GitHub, live API documentation, and through a static-website\footnote{\url{https://atlarge-research.github.io/opendt/}}, containing a demo, a tutorial~\ref{design:uc3:education} presented through a series of pre-recorded videos.

\section{Conclusion and future work}\label{sec:conclusion}

Understanding the performance and sustainability of operating and upcoming datacenters is essential to our society and economy. 
Addressing the lack of an open-source digital twin for datacenters, in this work we have designed, implemented, and experimented with \ourtool{}.  
Overall, our experimental results suggest that \ourtool{} supports continuous replay of real-world datacenter operations, with high accuracy and, through self-calibration, can improve this accuracy over state-of-the-art approaches.

We have released \ourtool{} as open-source and plan to conduct extensive future experimentation and coupled with real-world infrastructure as closed-loop, as part of a major infrastructure project with over 75 partner institutions~\cite{FutureNetworkServices2025}.
We plan to expand \ourtool{} to domain specific operation and couple closed-loop with large-scale ICT infrastructure, running LLM inference workloads. 
Lastly, we plan to build educational material around digital twinning, aided by \ourtool{}, and release as open-education, and as optional material in computer systems courses on Computer Organization, Distributed Systems, or in a future edition of the course on Modern Distributed Systems MOOC from edX~\cite{delftx_modern_distributed_systems}, which already uses a form of \oursim{} and could therefore use an exercise based on \ourtool{}.

\section*{Acknoweldgement}
This work is partially supported by 
EU MSCA CloudStars (101086248) and 
Horizon Graph Massivizer (101093202), and 
by the NL National Growth Fund 6G flagship project Future Network Services.

\bibliographystyle{ACM-Reference-Format}
\bibliography{main.bib}

\end{document}